\documentclass[acmsmall,screen,nonacm]{acmart}
\usepackage{array}

\AtBeginDocument{%
  
}

\setcopyright{none}
\newcolumntype{L}[1]{>{\raggedright\arraybackslash}p{#1}}

\acmYear{2026}
\copyrightyear{2026}
\begin{document}

\title[Visibility Asymmetry]{Visibility Asymmetry: How Vendor Attention Shapes Which EdTech Breakdowns Become Product-Visible}

%% Author metadata confirmed for submission: unaffiliated independent submission, location disclosed at state level only.
\author{Lucan Li}
\email{lilucan1127@gmail.com}
\affiliation{%
  \institution{Unaffiliated}
  \city{Indiana}
  \country{USA}
}

\renewcommand{\shortauthors}{Li}

\begin{abstract}
CSCW infrastructure scholarship often treats breakdown as the moment at which infrastructures become visible. But in vendor-managed sociotechnical systems, not all breakdowns become visible to actors with repair capacity. Drawing on a retrospective qualitative corpus from a Chinese K--12 EdTech deployment, including 11 interviews, 5 classroom observations, and 28+ days of field notes, this poster introduces \emph{visibility asymmetry}: a sensitizing concept for situations in which similar local breakdowns encounter uneven routing conditions. I trace a four-stage mechanism: procurement categories sort schools into attention tiers; staffing and visit cadence follow those tiers; only some local problems travel through staff or administrator channels; and dashboards can re-code unresolved repair labor as adoption. The concept extends CSCW work on infrastructure, articulation work, and repair by shifting attention from the occurrence of breakdown to the organizational channels through which breakdown becomes actionable. I discuss implications for feedback systems that surface repair without expanding surveillance.
\end{abstract}



\maketitle

\section{Introduction: Not Every Breakdown Travels}
\label{sec:intro}

A classroom technology can fail in ways that teachers know, on-site staff know, and students experience, while remaining invisible to the product team that could change it. In a vendor-managed K--12 EdTech deployment, a random-selection feature that wastes class time, a tablet-charging routine that requires daily staff labor, or a homework recommendation system that requires teacher re-verification may all appear in usage metrics as successful adoption. What differs is not whether breakdown occurs, but whether the breakdown travels: whether it is observed during a product visit, translated by on-site staff, escalated by an administrator, or absorbed locally as ordinary teacher work.

This poster introduces \emph{visibility asymmetry} to name this routing problem. Building on CSCW work on infrastructure, articulation work, and repair, I argue that breakdown visibility is not simply an epistemic moment produced by failure. It is an organizational achievement, unevenly distributed across vendor-attention tiers. The contribution is conceptual: visibility asymmetry offers a vocabulary for analyzing how vendor-managed infrastructures selectively convert local repair labor into product-visible problems. Empirically, the concept is grounded in a retrospective qualitative corpus from one Chinese K--12 EdTech vendor deployment.

The paper deliberately does not claim to estimate the rate at which breakdowns become visible across schools. The corpus was not designed for that kind of distributional proof. Instead, the poster offers visibility asymmetry as a \emph{sensitizing concept}: a portable analytic vocabulary for asking who can notice breakdown, who can translate it, and which organizational channels can turn local repair into system-level change.

\section{From Breakdown Visibility to Routing Visibility}
\label{sec:background}

CSCW and STS infrastructure scholarship has long shown that infrastructure is relational, embedded in practice, and often visible upon breakdown~\cite{starRuhleder1996,star1999ethnography}. Classification work further shows that infrastructures make some categories and forms of labor easier to see than others~\cite{bowkerStar1999}. Repair scholarship shifts attention from failure as exception to maintenance and repair as constitutive work~\cite{jackson2014}. Articulation-work scholarship similarly foregrounds the connective labor required to align distributed work so that cooperative activity can proceed~\cite{strauss1985,schmidtBannon1992,suchman2007,starStrauss1999}. Recent work further argues that as platforms have become embedded in institutional life, the boundary between platform and infrastructure has blurred, giving platform owners disproportionate control over what counts as a problem and what is recorded as use~\cite{plantin2018platform}.

These traditions are central to the present analysis, but vendor-managed educational infrastructure exposes an additional problem: breakdown may be locally visible and still organizationally invisible. A teacher can know that a feature fails, an on-site staff member can know that a workaround is routine, and a school administrator can know that usage is performative, while the product organization sees only adoption traces. Visibility asymmetry names this gap between local knowing and product-visible repair capacity.

The concept is therefore not a replacement for articulation work or for Star and Strauss's analysis of silencing. It inherits those concerns and specifies a different unit of analysis. Articulation work explains the coordination and translation labor that keeps distributed work moving; layers of silence explains how some work becomes muted or voiceable in organizational arrangements. Visibility asymmetry asks how local repair episodes become, or fail to become, product-side records in a vendor--school infrastructure. The unit shifts from voice and coordination alone to the inter-organizational routing topology that makes a breakdown actionable.

This gap matters for CSCW because deployed organizational technologies are rarely maintained by the same people who absorb their everyday failures. Teachers, students, school administrators, vendor staff, account managers, and product teams occupy different locations in the cooperative infrastructure. Their positions shape what they can see, what they can report, and what kind of problem formulation can travel upward. In this sense, breakdown visibility is not only a property of the artifact or the use situation. It is a property of a routing chain.

\section{Empirical Grounding and Ethical Boundaries}
\label{sec:grounding}

The concept is grounded in a retrospective qualitative corpus generated during employment with one Chinese educational-technology vendor: 11 interviews, 5 classroom observations, and 28+ days of field notes across eight focal schools and related site visits. The deployed suite included classroom-interaction, homework-workflow, and AI-innovation modules. Teachers were mandated end-users rather than voluntary adopters; on-site vendor staff supported training, hardware setup, printing and scanning cycles, and first-line troubleshooting.

For this poster, I do not attempt to summarize the full repair-labor analysis. Instead, I use the corpus to theorize one cross-cutting pattern: similar breakdown categories appeared across sites, but their likelihood of becoming product-visible varied with vendor attention, staffing, and routing channels. I use ``vendor-attention tiers'' as an analytic shorthand rather than as a neutral taxonomy. The highest-attention category is anchored in vendor-internal language around flagship or demonstration sites; the standard-staffed and thinly covered categories are researcher-constructed contrasts based on staff coverage, product-side visit cadence, and whether teacher concerns had plausible routes beyond the school layer. This mixed emic/etic vocabulary is analytically relevant: the vendor had clearer language for sites meant to be made visible than for sites where breakdowns were more likely to be locally absorbed.

Because the corpus contains non-public workplace artifacts and potentially identifying details, all school, region, module, and participant names are pseudonymized. High-risk staff passages are paraphrased rather than directly quoted. The author's own position within the vendor's deployment infrastructure shaped which breakdowns were observable and which remained out of frame; this analysis is therefore itself a partial routing of the deployment, not a view from outside it. The claim here is not that visibility asymmetry has been fully validated as a general model, but that it captures a recurring routing problem that CSCW research on organizational infrastructures should be able to name.

\section{Visibility Asymmetry}
\label{sec:concept}

I use \emph{visibility asymmetry} to name a structural condition in vendor-managed sociotechnical infrastructure: classroom-level breakdowns of similar form become legible to product repair channels under uneven routing conditions, depending on each site's position in a vendor-attention hierarchy. The asymmetry is not in whether breakdowns occur. It is in whether the local repair labor that absorbs them is routed upward into product-side problem records or absorbed silently into teacher and staff workflows.

The mechanism has four linked stages. First, procurement and conversion logic sort deployed sites into attention tiers. In this corpus, vendor attention was not simply an internal product decision; consistent with critical accounts of EdTech as a commercial sector in which vendor business logics, not only product logic, structure platform--school relationships~\cite{komljenovic2021rentiers}, it was shaped by procurement arrangements, demonstration value, and conversion-oriented site management, which made some schools more useful as reference sites and therefore more likely to receive staff and product-side attention. Second, staffing and product-visit cadence follow those tiers. Third, local breakdowns travel upward only where staff, administrator, or product-visit channels exist. Fourth, dashboards re-code the resulting distribution: high-attention sites may appear as sites of legible repair, while lower-attention sites may appear as sites of adoption even when adoption is sustained by local workaround labor.

\begin{figure}[t]
\centering
\includegraphics[width=\linewidth]{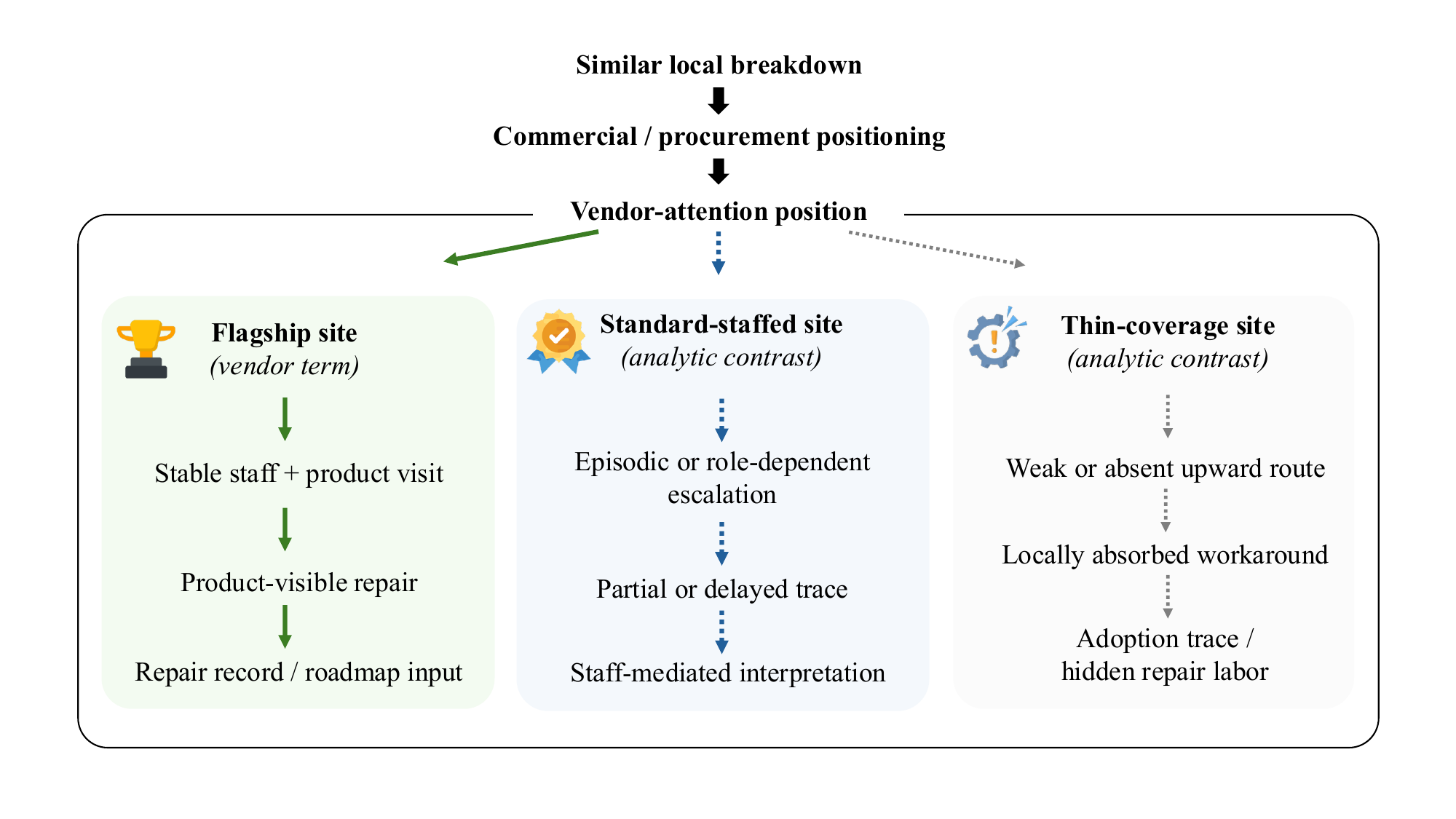}
\caption{Visibility asymmetry as a routing condition: the same local breakdown can become product-visible repair, a partial trace, or a locally absorbed workaround depending on vendor-attention position and available escalation channels.}
\Description{A color-coded routing diagram beginning with similar local breakdown and commercial or procurement positioning. It branches into three vendor-attention positions: flagship site as a vendor term, standard-staffed site as an analytic contrast, and thin-coverage site as an analytic contrast. Each branch shows different staffing and escalation routes leading to product-visible repair, partial or delayed trace, or locally absorbed workaround, and then to repair record, staff-mediated interpretation, or adoption trace with hidden repair labor.}
\label{fig:routing}
\end{figure}

Figure~\ref{fig:routing} summarizes the concept. Its purpose is not to replace empirical comparison with a deterministic model. Rather, it provides a vocabulary for asking a different CSCW question. Instead of asking only whether an infrastructure breaks down, we can ask: through which actors does breakdown become speakable, recordable, and actionable? This formulation also avoids making the thin-coverage limit case carry the concept by itself: the core phenomenon is not a three-tier distributional proof, but the routing condition under which any local breakdown, including one inside a flagship site, becomes product-visible rather than absorbed as use.

\section{Grounding Cases Across Attention Tiers}
\label{sec:cases}

The corpus offers three grounding cases that make the concept analytically useful while also showing its evidentiary limits. Table~\ref{tab:examples} summarizes representative breakdown categories and their differing routing contrasts across attention positions.

\subsection{High-attention sites: breakdowns with repair channels}

At high-attention sites, breakdown did not disappear; it became easier to attach to a repair channel. One teacher's difficulty recovering her own published assignments became a staff-mediated request for a ``my recent assignments'' view. In an observed primary classroom, QR-code scanning, low-resolution resources, random-selection failures, and tablet-holding burdens occurred inside a flagship deployment. These were still local frictions. The difference was that they were more likely to be observed, narrated, and routed because product-side visits and stable staff coverage made the site more legible to the vendor.

This case guards against a misleading interpretation of visibility asymmetry. High attention does not mean high functionality. It means that the breakdown has a better chance of being connected to organizational repair capacity. The same site can therefore be both a successful reference deployment and a place where substantial repair labor is performed.

\subsection{Standard-staffed sites: role-dependent escalation}

At standard-staffed sites, similar breakdowns surfaced, but upward routing was more episodic and role-dependent. One teacher described a points-system data-loss problem recurring several times before backend rollback, after she repeatedly escalated through vendor staff and insisted that the problem was vendor-side. In the homework module, sophisticated design needs traveled more plausibly through an administrator or subject-group head than through ordinary classroom use. The same kind of breakdown thus required more positional work to become product-visible.

The CSCW point is not simply that communication was difficult. It is that the communication channel itself was part of the infrastructure. A classroom teacher, a subject-group head, a vice principal, and an on-site vendor staff member can each formulate a problem differently and with different reach. Visibility asymmetry directs attention to these role positions and to the translation labor that connects them.

\subsection{Thin coverage as a limit case: local absorption}

At the thinly covered limit case, repair expanded into withdrawal and parallel infrastructure. A teacher maintained lesson preparation, supplementary materials, and interaction practices outside the vendor system, logging in mainly when administrative compliance required it. The platform remained dashboard-present but pedagogically peripheral. This is not simply non-adoption. It is a form of local absorption in which repair labor sustains the appearance of deployment while reducing the chance that breakdown will become product-visible.

This case should not carry the full evidentiary weight of a tier theory. It is a boundary case: useful because it shows what happens when routing channels are weakest, but insufficient for making general claims about all low-attention or under-resourced schools. For a poster-length concept paper, that limitation is acceptable if the claim remains conceptual rather than distributional.

\begin{table}[!htbp]
\centering
\caption{Grounding examples for visibility asymmetry.}
\label{tab:examples}
\footnotesize
\begin{tabular}{@{}L{3.1cm}L{4.25cm}L{3.65cm}@{}}
\toprule
Breakdown & Routing contrast & Conceptual use\\
\midrule
Random selection & Observed/logged during some high-attention visits; elsewhere requires teacher escalation & Visibility depends on available route\\
Tablet setup & Known as staff labor locally but absent from dashboard metrics & Repair is not automatically counted\\
Courseware quality & Observable in flagship use; elsewhere may produce parallel workflows or withdrawal & Adoption traces can hide substitution\\
Homework recommendation & Staff-mediated request in one site; role-based escalation in another & Repair depends on position\\
\bottomrule
\end{tabular}
\end{table}

\section{Implications for CSCW}
\label{sec:implications}

\textbf{Breakdown is not enough; routing matters.} For CSCW infrastructure studies, visibility asymmetry clarifies that breakdown does not automatically create visibility. Breakdown becomes analytically and organizationally consequential only when it travels through actors, records, and repair channels. The unit of analysis is therefore not only the local failure but the path by which that failure can become actionable.

\textbf{Dashboards can classify repair labor as adoption.} Learning-analytics and dashboard research has emphasized teacher-centered, situated, and co-designed dashboards~\cite{wiley2024teacherDashboards,sarmientoWise2022codesign}. Visibility asymmetry adds a routing question to this agenda. Platform use is not a transparent proxy for successful adoption. Usage traces may be produced by local workaround labor, administrative pressure, or staff mediation. A repair-informed dashboard would need to represent not only activity, but the labor required to make that activity possible.

\textbf{Surfacing repair must not become surveillance.} The design challenge is not simply to count more teacher labor. HCI research on hidden data work and downstream AI failures shows that making invisible labor visible can either support repair or create new burdens~\cite{sambasivan2021data}; recent educator-centered analyses of EdTech AI harms similarly caution that teacher experience must ground assessment of system effects rather than serve as a downstream metric~\cite{harvey2025teachers}. In school settings, logging every workaround risks turning repair into another performance metric, extending the platformed classroom surveillance that critical EdTech scholarship has documented in datafied school platforms~\cite{manolev2019datafication}. The CSCW problem is to design feedback systems that help local breakdowns travel to actors with repair capacity without converting every workaround into a surveillance trace.

\textbf{The concept can travel beyond EdTech.} Although grounded in K--12 EdTech, visibility asymmetry may apply to other vendor-managed infrastructures: health IT, enterprise software, platform labor systems, and public-sector digital services where local users and frontline staff absorb breakdowns that product teams never see. The concept is useful where the people who keep a system working are not the same people who decide what gets fixed.

\section{Limitations and Next Steps}
\label{sec:limits}

This poster makes a bounded conceptual claim. The corpus is retrospective and generated from industry fieldwork rather than prospective academic research. The evidence does not estimate visibility rates, compare all tiers systematically, or include product-team ticket data. The staff-side analysis relies heavily on one on-site staff account, and the thin-coverage evidence is a limit case. These are real limits, not minor caveats.

The next step is to test and refine visibility asymmetry across settings where repair channels can be observed more directly: support tickets, product-roadmap meetings, account-management records, teacher feedback tools, and administrator mediation. The broader research question is whether systems designed around adoption metrics can also support repair visibility without making teachers and frontline staff more surveilled.

\section*{Data Availability and AI Use Disclosure}

Raw data cannot be publicly shared because the corpus contains non-public workplace field artifacts and potentially identifying information about teachers, schools, vendor staff, and deployment sites. This poster provides anonymized and paraphrased examples sufficient to ground the concept.

Generative AI tools were used for language-level drafting support, consistency checks, and revision planning. The author retained responsibility for all claims, analysis, source selection, and reference verification.

\bibliographystyle{ACM-Reference-Format}
\bibliography{references}

\end{document}